\documentclass[12pt]{article}
\def\d#1/d#2{ {\partial #1\over\partial #2} }

\newcommand{\Red}{\ }
\newcommand{\Blue}{\ }

\newcommand{\m}[1]{ {\Red $#1$} }

\newcommand{\beq}{\Red \begin{eqnarray}}
\newcommand{\eeq}{\end{eqnarray}\Blue}
\newcommand{\numeq}{\end{eqnarray}\Blue}

\newcommand{\half}{ {1\over 2} }

\newcommand{\pdr}{\partial}

\usepackage{graphics}
\input{epsf}

\newcommand{\beqs}{\begin{eqnarray}}
\newcommand{\eeqs}{\end{eqnarray}}

\def\m#1{$#1$}
\newcommand{\Em}{\em}

\def\tr{\;{\rm tr}\;}

\def\mod{\;{\rm mod}\;}

\begin{document}

\centerline{\bf\large New Classical Limits of Quantum  Theories\footnote{To Appear in the Proceedings of the 
70th  Meeting of Mathematicians and Physicists at   Strassbourg, June 2002; ed. by V. Turaev and T.
Wurzbacher.}
}

\begin{center}{ S.~ G.~ Rajeev\footnote{rajeev@pas.rochester.edu}\\
    Department of Physics and Astronomy,\\
    University of Rochester,\\
    Rochester, New York 14627 \\
   }
   \end{center}
   
\abstract{ Quantum fluctuations of some systems vanish not only in the limit \m{\hbar\to 0},  but  also as some
other parameters (such as \m{1\over N}, the inverse of the number of `colors' of a Yang-Mills theory) vanish. These lead to 
new classical limits  that are often much better approximations to the quantum theory. 
We describe two examples: the familiar Hartree--Fock-Thomas-Fermi methods of atomic physics as well as the
limit of large spatial  dimension. Then we
present an approach  of the Hecke operators on  modular forms inspired by these ideas of
 quantum mechanics. It explains in a simple way why the spectra of these operators tend to the spectrum of
 random matrices for large weight for the modular forms.
}
   \vfill\eject
   \tableofcontents
  \vfill\eject
   
   \section{Introduction}
   
   It is well-known that a classical mechanical  system has many possible
   quantizations.  The classical theory is the limiting case as \m{\hbar\to 0}
   , so it is not surprising that there would be many quantum theories that
   in this limit reduce to the same classical theory. In this, largely expository, paper I will
   discuss the opposite phenomenon: how the same quantum theory can be 
   obtained by quantizing radically different classical systems. Viewed another
   way, a quantum theory could depend on two  parameters, say
   \m{\hbar,\lambda} and the quantum fluctuations of  some class of observables are of order
   \m{\hbar\lambda}..
    Then  both  the limits \m{\hbar\to 0} and \m{\lambda\to
   0} are classical theories. These classical theories could be entirely different.
   In an example from 
   atomic physics,  the conventional classical limit 
   has  a finite number of degrees of freedom, while the new one  has an 
   infinite number.  We will refer to the   the new limits (obtained
   by taking parameters other than \m{\hbar} to zero)  as 
   `neo-classical limits'.
   
   This phenomenon is physically interesting because  one of
   the new classical limits may be a better approximation to the quantum theory than the naive classical
   limit. These ideas came into the  high energy physics literature from the work of 't Hooft and Witten on the
   large N limit of gauge theories. Witten \cite{witten} in particular worked out several simpler cases to 
   popularize the
   notion that even case \m{N=3}  may be well approximated by the large N limit. But historically, 
   the various mean field theories of condensed matter physics \cite{largeN} (the spherical
   model for example) and even the theory of Nuclear Magnetic Resonance can be thought of as precursors of
   these ideas.

   For example, in atomic physics,  in the usual classical
   limit \m{\hbar\to 0}  there is  no ground state: the hamiltonian is not bounded
   from below. However, the  neo-classical limit (in this case a version  of the
   Hartree-Fock approximation) has a ground state. Moreover, it gives an
   excellent first approximation to the ground state energy of the atom. The
   neo-semiclassical expansion gives a systematic way of calculating corrections
   to arbitrary accuracy, although the complexity grows rapidly with the desired
   accuracy.

   Another  important example, discussed below, is also from atomic physics: the quantum
   fluctuations in electron distances become small in the limit as the dimension of space becomes large. 
   We will derive a simple effective potential  that explains the stability of
   the atom. For another approach to this see the work of Hershbach \cite{hershbach}.
   
   Some important corners of mathematics are also  illuminated  by this phenomenon. The theory of
   modular forms can be viewed as the quantization of a classical mechanical system whose phase space is the
   upper half plane. The limit as the weight of the modular form goes to infinity corresponds to a classical
   limit. But there is also another classical limit corresponding to letting the level (the area of the
   fundamental domain ) go to infinity. These limits lead to interesting new approaches to the problem of
   determining the spectrum of the Hecke operators on modular forms.
   
   The mathematical formulation  of  a classical dynamical system has expanded steadily in generality 
   throughout history: as new physical
   theories are discovered we are led to enlarge the formalism to incorporate  the new developments. In the
   progression from ordinary differential equations to  Hamilton-Jacobi theory, symplectic geometry and
   the currently fashionable Poisson algebra formulation, we learn to deal with increasingly
   sophisticated systems and symmetries. Neo-classical limits produce classical systems of even greater generality, often
   with non-local action principles and no simple hamiltonian description \cite{3dconf}. We have described some examples 
    of this before.   There are classical limits of quantum field theories that retain
    asymptotically freedom and require a  renormalization \cite{2dthirring}.
    The   challenge of finding the right mathematical description of these 
    new kinds of classical systems remains. 
   
   Much of the story  told  in this paper is, of course, well-known. I hope that organizing them
   in this way will help to understand common themes in apparently distant subjects.

   \section{Hartree--Fock Theory of Atoms}
   
   \subsection{The Classical Limit of the Atom}
   We start with a basic problem of quantum mechanics that cannot be solved exactly: an atom (or ion) with  more
   than one electron. We usually start with the classical hamiltonian
   \beq
   H=\sum_{i=1}^m {p_i^2\over 2\mu}-\sum_{i=1}^m{Ze^2\over |r_i|}+\sum_{1\leq i<j\leq m}{e^2\over |r_i-r_j|}.
   \eeq
   As usual, \m{Z} is the atomic number of the nucleus, \m{e} the charge of the electron and \m{\mu} its mass.
   For each \m{1\leq i\leq m}, the position \m{r_i} and momentum \m{p_i} of the electron are vectors in
   the Euclidean space \m{R^3}.
   Then we pass to the quantum theory whose hamiltonian is
   \beq
   \hat H=-\hbar^2\sum_{i=1}^m {\nabla_i^2\over 2\mu}-\sum_{i=1}^m{Ze^2\over |r_i|}+\sum_{1\leq i<j\leq m}
   {e^2\over |r_i-r_j|}.
   \eeq
   This is an operator on the complex Hilbert space of anti-symmetric wavefunctions 
   \m{{\cal F}_m=\Lambda^m\left({\cal H}\right)}. The anti-symmetry 
    incorporates the Pauli exclusion principle. The space of 
   single particle wavefunctions is \m{{\cal H}=L^2(R^3, C^2)};  the wavefunction of each electron
   takes values in \m{C^2}, since it  can exist in two spin states
   \footnote{ We will, for simplicity, ignore relativistic and spin-dependent terms in the hamiltonian.}.
   
   The limit \m{\hbar\to 0} is the usual classical limit of the atom. It is well-known that this is a
   spectacularly bad approximation  to the quantum theory of the atom. The quantum hamiltonian is
   self-adjoint and bounded below and hence has a well-defined ground state. Indeed the central problem of
   atomic physics is the determination of this ground state wavefunction  and the corresponding eigenvalue.
   The hamiltonian of the classical limit on the other hand,  has no ground state: we can  let  position of
   an electron approach the nucleus, \m{r_a\to 0}, thus descreasing the energy down to
   \m{-\infty}. 
   
   What is missing here is  the uncertainty principle: in the quantum theory it is not
   possible to make the position of the electron close to the nucleus without making its kinetic energy
   large. This might suggest that there is no way to produce a classical approximation to the atom with a
   stable ground state. 
   
   We will now produce a completely different classical system (with an infinite number of degrees of
   freedom in fact) whose quantization yields exactly the above 
   quantum theory of the atom. Moreover, it has a ground state which is even a good approximation to the
   quantum ground state. What  we will describe is just a reformulation of  the standard
   Hartree-Fock approximation in atomic physics.  This reformulation allows  a generalization to relativistic
   many-fermion systems which we have described elsewhere \cite{2dhadron}.
   
   \subsection{ The Neo-Classical Theory of the Atom}
   \def\Gr{{\rm Gr}_m\left({\cal H}\right)}
   There are many standard texts that discuss the material in this section, although often  without the geometric
   interpretation in terms of the Grassmannian. See for example \cite{hartreefock}
   Let \m{{\cal H}=L^2(R^3,C^2)} be the familiar complex Hilbert space. We define the {\Em Grassmannian} 
    \m{\Gr} to be the space of linear self-adjoint trace-class\footnote{ If a projection operator is trace class, its trace must be an integer, the dimension of the
    vector space it projects.} projection operators of rank \m{m}:
    \beq
    \Gr=\{\rho:{\cal H}\to {\cal H}| \rho^\dag =\rho;\rho^2=\rho;\tr\rho=m\}.
    \eeq
    It is clear that an eigenvalue of \m{\rho} is equal  either to zero or to one.
   Corresponding to each such projection operator there is  a subspace of \m{\cal H} of dimension \m{m}: the
   eigenspace of \m{\rho} with eigenvalue one. Conversely, each such \m{m}-dimensional subspace \m{V}  
   defines an orthogonal decomposition \m{{\cal H}=V\oplus V^\perp}; then we can construct  \m{\rho} as 
   the  hermitean projection operator to \m{V}. Thus we can see that \m{\Gr} is really the set of all
   \m{m}-dimensional subspaces of  \m{\cal H}. Thus we have an infinite dimensional (but finite rank) 
   generalization of the  usual definition of the Grassmannian \cite{chern}. \m{\Gr} is an infinite
   dimensional manifold, whose tangent space is the space of rank \m{m} self-adjoint operators on \m{\cal H}.
   
   A classical dynamical system is specified by  (i) a manifold which  will be its
   phase space, (ii) a symplectic form on this phase space which will determine the Poisson brackets, 
   and (iii) a  real function on the phase space which is its hamiltonian. 
   
   In our theory, \m{\Gr} is the phase space. The symplectic form on it generalizes the standard symplectic
   form on finite-dimensional Grassmannians. The Poisson brackets that
   it implies for a pair of functions is:
   \beq
   \{f,g\}=\tr \rho [df,dg].
   \eeq
   Here \m{df} is the infinitesimal variation of \m{f} which can be thought of as a linear operator on 
   \m{\cal H}.
   If we represent the operator \m{\rho}  by its integral kernel \m{\rho^a_b(x,y)} ( where \m{a=1,2} labels
   spin) we can write this Poisson bracket as
   \beq
   \{\rho^a_b(x,y),\rho^c_d(z,u)\}=\delta^c_b\delta(y,z)\rho^a_d(x,u)-\delta^a_d\delta(x,u)\rho^c_b(z,y).
   \eeq
   
    The last piece of information is the hamiltonian, which we postulate to be
   \beqs
   H_1&=&\int {p^2\over 2\mu}\tilde\rho(x,p) d^3x {d^3p\over (2\pi\hbar)^3}
   -\int{Ze^2\over |x|}\rho^a_a(x,x) d^3x\cr 
   & & +\half\int{e^2\over |x-y|}\left[\rho^a_a(x,x)\rho^b_b(y,y)-\rho^a_b(x,y)\rho^b_a(y,x)\right]d^3xd^3y.
   \eeqs
   Here, \m{\tilde\rho} is the {\em symbol} of the operator \m{\rho}: 
   \beqs
   \tilde\rho^a_b(x,p)&=&\int \rho^a_b(x+{u\over 2},x-{u\over 2})e^{-{i\over
   \hbar}p\cdot u}d^3u,\cr 
   \rho^a_b(x,y)&=&\int \tilde \rho^a_b({x+y\over 2},p)e^{{i\over \hbar}p\cdot (x-y)}{d^3p\over
   (2\pi\hbar)^3}.
   \eeqs
    Clearly, 
   \m{\rho^a_a(x,x)=\int \tilde\rho^a_a(x,p){d^3p\over
   (2\pi\hbar)^3}}.
   
   So far it is clear that this  system depends on the parameters \m{\mu,e,Z,m} of the 
   of the atom. Although the dynamical variables are operators, and \m{\hbar} appears in the formula for the
   hamiltonian, it is a bona fide classical dynamical system. 
   
   We will  show that a  quantization of this system is exactly the quantum theory of the atom. The
   physical meaning of the operator \m{\rho} is that it is the `density matrix'  of the electrons. Indeed
   \m{\rho^a_a(x,x)} is the number density of the electrons at the point \m{x};
   \m{\tr\rho=\int \rho^a_a(x,x) d^3x=m} is the total number of electrons.
   More generally, \m{\tilde\rho^a_a(x,p)} is the
   density of electrons of momentum \m{p} and position \m{x}. The Pauli exclusion principle which allows for
   at most one electron per single particle state, becomes the condition that this density matrix be a 
   projection operator, so that its eigenvalues can only be zero or one. Thus this classical dynamical system
   realizes many of the facts we usually associate with quantum theory.

   The first term represents the kinetic energy and  the second term the potential energy due to the nucleus.
   We can combine these `single-particle' terms in the hamiltonian  into the form
   \beq
   \tr \rho K,\quad  K=-{\hbar^2\over 2\mu}\nabla^2-{Ze^2\over |x|}.
   \eeq
   
   Now, \m{K} is bounded below by \m{E_1=-\half \mu{Z^2e^4\over \hbar^2}}, as we know from the elementary theory 
   of an ion with one electron. Since \m{\rho} is positive and \m{\tr \rho=m}, we see that \m{\tr\rho K\geq
   mE_1}. ( A stricter bound  can be obtained using the fact that \m{\rho} is a projection.
   But we don't need it.) This way,   the system avoids the catastrophe of the conventional classical limit.
   \footnote{Strictly speaking \m{H_1} exists only on some dense domain of \m{\Gr}. This
   domain should be some class of pseudo-differential operators, and  is the true phase space of the system. The
   correct statement is that the hamiltonian is bounded below within this domain. This
   is a  technical  project that I am unable to complete. It
   would be interesting to produce a  functional analytic realization of the physical ideas that are
   described here.}.

    The
   interaction of the electrons induces two kinds of terms. The first is obvious, the Coulomb energy of a
   charge cloud of density \m{e\rho^a_a(x,x)}. The  last term is not so obvious--it is the `exchange energy'. It
   is needed to get back the correct quantum theory (see below). By a version of the Schwarz inequality it
   should be possible to see that 
   \beq
   \int\left[\rho^a_a(x,x)\rho^b_b(y,y)-\rho^a_b(x,y)\rho^b_a(y,x)\right]{e^2\over |x-y|} d^3x
   d^3y\geq 0.
   \eeq
    This expresses the physical fact that the electron-electron interaction is repulsive.
     Thus the total hamiltonian is bounded below by at least \m{mE_1}.
   
   To actually find the ground state of this classical system, we must vary the hamiltonian subject to the
   constraints on \m{\rho}. Such a variation of \m{\rho} is always of the form \m{\delta\rho=-i[\rho, u]} for
   some hermitean operator \m{u}. The condition for an extremum is then 
   \beq
   	[\rho, dH_1]=0.
   \eeq
   Here \m{dH_1={\pdr H_1\over \pdr \rho}} is a linear operator
   \beq
   dH_1=K+{\cal U}+{\cal W}.
   \eeq
   Here, \m{K} is as defined above and 
    \m{\cal U} is the multiplication  by the `mean field'
   \beq
   {\cal U}^a_b(x)=\delta^a_b\int  {Ze^2\over |x-y|}\rho^a_a(y,y)d^3y.
   \eeq
   The `exchange energy' contributes an operator \m{W} whose integral kernel is
   \beq
   {\cal W}^a_b(x,y)=-{Ze^2\over |x-y|}\rho^a_b(x,y).
   \eeq
   
   Thus we can find the extremum  by solving a non-linear eigenvalue problem self-consistently. We have,
   \beq
   \rho=\sum_{a=1}^m \psi_a\otimes \psi^\dag_a
   \eeq
   where each of the vectors \m{\psi_a\in {\cal H}} is an eigenstate of \m{dH_1}.  
   
   This is exactly the Hartree-Fock approximation to the atomic ground state. 
   We find the  wave-functions that are eigenstates of some single particle hamiltonian; the potential in
   this hamiltonian is self-consistently determined by postulating that \m{m} of these are occupied by
   electrons. Our description avoids the usual Slater determinants for the wavefunction-the hamiltonian 
   only depends on the
   density matrix of the electrons and not the wavefunction itself. We have shown that this way of
   formulating the Hartree--Fock theory allows for generalization to systems containing an infinite number of
   fermions such as relativistic theories \cite{2dhadron,2dthirring}. 
   
   Our point in this paper is that this
   theory  can be thought of as minimizing the hamiltonian of a classical system on \m{\Gr}.
   This extends to  the time evolution as well: the hamiltons equations of
   our system are the usual equations of time-dependent Hartree-Fock theory.
   
   \subsection{ Back To the Quantum Theory}
\def\fGr{{\rm Gr}_m\left({\cal V}\right)}
 
   How do we quantize a system whose phase space is a Grassmannian? It is not possible to cover the
   Grassmannian by a single co-ordinate system, so it is inconvenient to look for canonical variables. 
   However, the Grassmannian is a K\"ahler manifold, and we can apply the ideas of geometric (or
   Berezin-Toeplitz) quantization \cite{kahlerquantization}. In an earlier paper (the appendix of \cite{2dhadron}) we used the 
   representation theory of the unitary group to quantize this theory.
   
   Recall the situation in the case of finite dimensional Grassmannians: let  \m{{\cal V}} be a finite
   dimensional vector space, and \m{\fGr} the set of its \m{m}-dimensional subspaces. \m{\fGr} is a compact
   K\"ahler manifold. Its canonical line bundle \m{{\cal L}} admits a hermitean metric and a connection whose curvature is
   just the symplectic form. ( A line bundle that admits such a metric and connection is said to be
   quantizable \cite{kahlerquantization}.) The holomorphic sections \m{{\rm Hol}({\cal L})} of this line bundle form a finite dimensional vector
   space isomorphic to \m{\Lambda^m\left({\cal V}\right)}. This space of holomorphic sections is a
   subspace of the Hilbert space of square-integrable sections with a projection operator  \m{\Pi:L^2({\cal L})\to {\rm Hol}({\cal L})}. These geometric facts can be used to
   construct a quantization \cite{kahlerquantization} of the dynamical system whose phase space is \m{\fGr}.
   
    From any function \m{f:\fGr\to
   R}  we will construct an operator \m{\hat f:{\rm Hol}({\cal L})\to {\rm Hol}({\cal L})} by the formula
   \beq
   	\hat f=\Pi f.
   \eeq
   That is, we multiply a holomorphic section by the function to get a section of \m{\cal L} that may not be
   holomorphic; then we simply project out the holomorphic part. The operator we construct this way is
   self-adjoint (it is just a finite dimensional hermitean matrix in fact).
   
   In what sense is it a quantization of the dynamical system on \m{\fGr}? How will we recover the classical
   limit? The idea \cite{kahlerquantization} is that it is merely a special case of a one-parameter family of
   quantum theories where  \m{{\cal L}} above is replaced by \m{{\cal L}^N}. As long as \m{N} is a positive
   integer the above ideas go through: there is still a projection \m{\Pi_N:L^2({\cal L}^N)\to {\rm Hol}({\cal
   L}^N)} to a finite dimensional space of holomorphic sections.
   \footnote{ The space \m{{\rm Hol}({\cal L}^N)} carries a representation of the
   unitary group \m{U({\cal V})} given by the Young diagram of height \m{n} and width \m{N}, generalizing the
   completely anti-symmetric tensor representation of the case \m{N=1} above.} Also in the limit \m{N\to
   \infty} the operator algebra tends to the Poisson algebra of functions in the sense that 
   \beq
  ||\Pi_Nf\ \Pi_Ng-\Pi_N(fg)||={\rm O}({1\over N}),
   \eeq
   and moreover
   \beq
   ||iN[\Pi_N f,\ \Pi_Ng]-\Pi_N(\{f,g\})||={\rm O}({1\over N}).
   \eeq
   Also the operator norm of \m{T_Nf} approaches the sup norm of the function \m{f}. 
   
   These ideas also extend to \m{\Gr} when \m{\cal V} is replaced by the infinite dimensional vector
   space \m{\cal H}. The technical aspects are simpler than in \cite{mickelssonrajeev,rajeevturgutcmp}, since we need only
   finite rank projections. We give only a very brief outline here. Any subspace of dimension  \m{m} can be brought to some standard subspace 
   whose orthogonal complement is \m{{\cal H}_\perp}; hence  \m{\Gr} is a coset space\footnote{\m{U_0({\cal
   H})} is the group  of unitary transformations that mixes the standard \m{m}-dimensional subspace with its
   orthogonal complement only by a  finite rank operator. It is analogous to the `restricted Grassmannian' of
   Sato \cite{PressleySegal} except that it is modelled on finite rank operators rather than compact ones.}
   \m{\Gr=U_0({\cal H})/U({\cal H}_\perp)\times U(m)}. Using the trivial representation of 
   \m{U({\cal H}_\perp)}
   and the determinant representation of \m{U(m)}, we can construct a line bundle 
   \beq
   {\cal L}=\left(U_0({\cal H})\times C\right)/U({\cal H}_\perp)\times U(m).
   \eeq
   The holomorphic sections of this bundle can now be constructed and shown to form \m{\Lambda^m({\cal H})}.
  It is thus clear that \m{\Lambda^m({\cal H})} is the  Hilbert space of at least one way of quantizing our
  system on \m{\Gr}. Indeed, the hamiltonian of the system when worked out in this way is exactly the 
  quantum hamiltonian of the atom we had earlier.

 This quantum hamiltonian is the special case as \m{N=1} of a one-parameter family of theories. For 
 \m{N>1} these describe fermions that carry a `color' quantum number, except that only observables that are
 invariant under \m{U(N)} are realized in the Hilbert space \m{{\rm Hol}({\cal L}^N)}. The Hartree--Fock
 method  thus approximates the theory for \m{N=1}  by the neoclassical limit as \m{N\to \infty}. In effect
 \m{1\over N} measures the size of the quantum corrections.
   
   A relativistic generalization of this theory is described in \cite{2dhadron}.
   There I developed an approach to two-dimensional QCD  where the large \m{N} limit was realized as a classical
   theory. The story above appears in the appendix to that paper. Later I found that many other problems in
   physics and mathematics can be thought of in a unifying way as different classical limits of the same
   quantum theory. 
   
   \section{ The  Thomas--Fermi Approximation }

The problem of minimizing the energy on the Grassmannian is still a hard problem. Further approximations are
needed. It turns out that there is a way to consider the limit\footnote{It is important to take 
\m{\hbar\to 0}, after the theory has been formulated on the Grassmannian as above; if
we took \m{\hbar\to 0} we would get a theory without a ground state, as explained above.}  \m{\hbar\to 0} (a kind of semi-classical
approximation) which yields a simpler theory. The ideas go back to the Thomas-Fermi approximation of early
atomic physics \cite{landaulifshitz} and have seen several revivals.  
There seems to be a connection with the density functional \cite{densityfunctional} method as well. Our point of view is based on symbol calculus and was in part
inspired by the work of Lieb, Thirring \cite{lieb} and others on the stability of matter. We will work out explicitly the leading terms but
indicate how higher order terms can be calculated systematically if needed.

It is possible to develop  the theory in a more general context than in the last section without much
additional work\footnote{In fact I worked this theory in this way (in 1992) to find  a relativistic generalization of the
Thomas-Fermi method. It remains unpublished.}. We therefore consider a system of $m$ fermions with the hamiltonian 
\beq
	H=\sum_{i=1}^m [T(-i\pdr_i)+U(r_i)]+\sum_{i< j}G(r_i,r_j).
\eeq
The configuration space of each fermion is \m{R^n}.
  We will allow the fermions to carry a `spin'
quantum number $\sigma=1,\cdots N_f$. The above hamiltonian is assumed to be
independent of this quantum number. (We will usually suppress the spin index).
Here, $T(p)$ is the `dispersion relation'; i.e., the dependence of kinetic energy on
momentum.$T(p)$  is usually spherically symmetric. $U(x)$ is the
external potential that all the fermions are subject to and $G(x,y)=G(y,x)$ is
the two body potential.

  In the last section  we had the following special case: the dimension of space \m{n=3}, the spin takes takes two
  values, $N_f=2$, the kinetic energy is $T(p)={p^2\over 2\mu}$ and the inter-electron potential is the
  Coulomb potential
  $G(x,y)={e^2\over |x-y|}$. The cases $n=1,2,3$ for the dimension of space are also of interest in other
  contexts.
  
  \def\Tr{\;{\rm Tr}}
  
  We have then the  Hartree--Fock energy
\beqs
     H_1(\rho)&=& \Tr (T+U)\rho \ + \ \cr
& & \half\int dxdyG(x,y)[\tr \rho(x,x)\tr \rho(y,y)-\tr \rho(x,y)\rho(y,x)]
\eeqs
to be minimized over all  operators satisfying  $\rho^{\dag}=\rho,\ \rho^2=\rho,\Tr \rho=m$. We
denote by $\tr$ the trace over flavor while $\Tr$  includes the integral over
position as well. As before, the first term represents the single--particle kinetic
energy and potential energy, the second term the direct interaction and the
last term the exchange interaction.

The minimization problem above leads to the variational equations
\beq
	[\rho,dH_{1}]=0
\eeq
where the Hartree--Fock self consistent hamiltonian itself depends on \m{\rho}.
A self--consistent solution is clearly\footnote{The theta-function of a self-adjoint operator
 \m{\Theta(A)} is defined to be the projection operator to the subspace on which \m{A} is positive.}
$\rho=\Theta(E_F-H_{1})$ where the `Fermi energy' 
$E_F$ is determined by the
condition $\Tr \rho=m$. It is often too hard to solve this  problem, so yet
another approximation is needed. We could minimize over some smaller set of operators \m{\rho} thereby
 obtaining a  variational bound that is simpler to calculate. Or we could calculate the function
 \m{H_1(\rho)} in a semi-classical approximation.

 The essence of the Thomas--Fermi approximation (a modern version is 
the density functional method \cite{densityfunctional}) is  a combination of these two ideas:\hfill\break
1. use the variational ansatz $\rho=\Theta(-h)$ where $h=t(-i\pdr)+v(x)$ is a
separable hamiltonian (i.e., a function of $p$ alone plus a function of $x$
alone);\hfill\break
2. expand  the energy function
\beq
	H_1(\Theta(-t-v))
\eeq
 semiclassically\hfill\break
3. minimize the leading term in this expansion $H_{TF}(t,v)$ with respect to the variational
 parameters $t$ and $v$. (
In many treatments, however, $t$ is chosen to be the same as $T(p)$ and only
$v(x)$ is varied.)

The semiclassical expansion will amount to an expansion in powers of the
derivatives of $v$. The above ansatz for $h$ is motivated by the form of the
Hartree--Fock  hamiltonian. If the two body potential $G$ is absent, the first
step is automatic, since $H_{1}$ is already in this form. Even for interacting fermions,  the direct 
interaction is already of the separable form.  The indirect
energy may not be separable in general, but as long as it is a monotonic
function of momentum, the projection operator $\Theta(E_F-H_{1})$ will agree
with that of some separable hamiltonian. Thus one expects this separable ansatz
 to be a good approximation.

The projection operator \m{\Theta(h-E)} can be expressed in terms 
 the resolvent operator, ${1\over h-E}$, since
 \beq
	\Theta(-x)=\int_D {dE\over 2\pi i} {1\over x-E}
\eeq where $D$ is a contour that surrounds the negative real axis in a
counterclockwise direction.
 There is a semi-classical expansion for the resolvent 
 which can be used to derive one for the projection operator
$\Theta(h-E)$.

\subsection{ The Semiclassical Expansion of the Resolvent}

To do the semiclassical expansion, it is convenient to restate the problem in
terms of  (matrix--valued) functions on the phase space $R^n\oplus R^n$ rather
than operators on the Hilbert space $L^2(R^n,C^{N_f})$.
 There is a systematic theory of this procedure (symbol
calculus) decribed in detail in for example, Ref. \cite{symbolcalculus}. The main idea
is to use Weyl ordering to set up a one--one correspondence
between functions on the phase space and operators on the
Hilbert space. From  a function $\tilde A(x,p)$ we construct
the operator $A$ whose kernel is
\beq
     A(x,y)=\int \tilde A({x+y\over 2},p)e^{{i\over \hbar} p\cdot (x-
y)}[dp].
\eeq
 We use the abbreviation $[dp]={dp\over (2\pi \hbar)^n}$.
If we apply this to simple functions such as polynomials we
can check that this definition corresponds to Weyl ordering. For example, the
function $\ xp\ $ becomes the operator $-i\hbar\pdr x +x (-i\hbar\pdr)$.

Conversely, given an operator, we define its symbol to be the
function
\beq
     \tilde A(x,p)=\int A(x+{z\over 2},x-{z\over 2})e^{-{i\over \hbar}p\cdot z}
dz.
\eeq
The idea is that $p$ is the momentum conjugate to the
{\it relative} coordinate of the operator kernel. The operator
multiplication can now be translated into the multiplication
of symbols. The result can be expressed in closed form:
\beq
     \tilde A\circ \tilde B(x,p)=\left\{
e^{{-i\hbar\over 2}(\d/d{x^i} \d/d{p_{i}^{'}}-\d/d{p_i} \d/d{x^{i'}})} \tilde
A(x,p)\tilde B(x',p')\right\}_{x=x';p=p'}.
\eeq
The trace of operators becomes an integral in phase space
\beq
	\Tr A=\tr\int A(x,x) dx=\tr\int dx[dp] \tilde A(x,p).
\eeq
We emphasize that the algebra of symbols under the multiplication law is exactly
the same (isomorphic) to the algebra of operators on a Hilbert space;i.e., no
approximation is involved in replacing an operator by its symbol.

We see that to the leading order the above muliplication law  is just the
pointwise
multiplication of the classical theory. In the next order
there is a correction proportional to the Poisson bracket.  If we expand the
exponential,
\beq
\tilde A\circ \tilde B(x,p)=\sum_{n=0}^{\infty}
\left({-i{\hbar}\over 2}\right)^n {1\over n!} \{\tilde A,\tilde B\}_{(n)}.
\eeq
Here we see a sequence of generalized Poisson brackets
\beq
	\{\tilde A,\tilde B\}_{(n)}=\sum_{r=0}^n(-1)^r
\tilde A^{j_1\cdots j_r}_{i_1\cdots i_{n-r}}
\tilde B^{i_1\cdots i_{n-r}}_{j_1\cdots j_r}
\eeq
where $\tilde A^i={\pdr \tilde A\over \pdr p_{i}}$ and
$\tilde A_i={\pdr \tilde A\over \pdr x^{i}}$ etc. $n=1$ corresponds to the
usual Poisson bracket. If  $\tilde A$ and $\tilde B$ commute as matrices on
spin, the odd brackets are antisymmetric and the even ones are
symmetric. Otherwise, there is no particular symmetry property.

Consider now the resolvent operator of a hamiltonian $h$,
\beq
	r(E)={1\over h-E}.
\eeq
We will now derive a semiclassical expansion for the symbol $\tilde r(E)$ of
this operator. The resolvent symbol  satisfies
\beq
	\tilde r(E)\circ (\tilde h-E)=1.
\eeq
Expand $r(E)$ in power series in ${\hbar}$ and put into the expansion of the
above equation to get,
\beq
	\tilde r(E)=\sum_{k=0}^{\infty} \tilde r_{(k)}(E) {\hbar}^k,
\eeq
\beq
	\sum_{n=0}^{\infty}\sum_{k=0}^{\infty}\left({-i{\hbar}\over 2}\right)^n{1\over
n!} {\hbar}^k\{\tilde r_{(k)}(E),\tilde h-E\}_{(n)}=1.
\eeq
Equating the powers of ${\hbar}$ on both sides of this equation, we get a set
of recursion relations
\beq
	\tilde  r_{(0)}(E)=(\tilde h-E)^{-1},
\eeq
\beq
	\tilde r_{(m)}(E)=-\sum_{n=1}^m({-i\over 2})^n{1\over n!}
\{\tilde r_{(n-m)}(E), \tilde h\}_{(n)} (\tilde h-E)^{-1}.
\eeq
If $\tilde h$ is diagonal in flavor space, the odd terms $\tilde r_{(2m+1)} $
vanish.  The above expansion can be used to derive the usual WKB quantization
conditions as well as higher order corrections to it.

Of particular interest to us is the case where $h$ is a separable operator:
\beq
	\tilde h(x,p)=t(p)+v(x).
\eeq
In this case the mixed derivatives in the generalized Poisson brackets vanish
and we get
\beq
\{\tilde r_{(k)}, \tilde h\}_{(n)}=\tilde r_{(k)i_1\cdots i_n}t^{i_1\cdots
i_n}+(-1)^n \tilde r_{(k)}^{i_1\cdots i_n}v_{i_1\cdots i_n}.
\eeq

If\footnote{ We use  units here such that \m{2\mu=1}, to simplify the formulas.}  $t(p)=p_ip_i$ as in nonrelativistic quantum mechanics, there is only one
term  for $n>2$,
\beq
\{\tilde r_{(k)}, \tilde h\}_{(n)}=(-1)^n \tilde r_{(k)}^{i_1\cdots
i_n}v_{i_1\cdots i_n}
\eeq
while
\beq
	\{\tilde r_{(k)}, \tilde h\}_{(1)}=2p_i\tilde r_{(k)i}-\tilde r_{(k)}^iv_i
\eeq
and
\beq
	\{\tilde r_{(k)}, \tilde h\}_{(2)}=2\tilde r_{(k)ii}+\tilde
r_{(k)}^{ij}v_{ij}.
\eeq
If moreover, $v$ is diagonal in flavor space, $\tilde r_{(1)}=0$ and
\beq
	\tilde r_{(2)}(E)={1\over 2(\tilde h-E)^2}[
{v_iv_i\over {\tilde h-E}}+2(p_iv_i)^2-v_{ii}].
\eeq
 
 \subsection{ Derivative Expansion of Energy Function}

Now we can rewrite the Hartree--Fock energy in terms of the
symbol $\tilde \rho(x,p)$ of the projection operator $\rho$.
\beqs
     H_1(\tilde \rho)&=& \tr \int \tilde \rho(x,p)T(p) dx[dp] +\tr\int
U(x)\rho(x) dx+\cr
& &\half\int G(x,y)\tr\rho(x)\tr\rho(y) dxdy\cr
& & -
{1\over 8}\int dx\int [dp][dp'] \tilde G(x,p-p')
\tr \tilde \rho(x,p) \tilde \rho(x,p')
\eeqs

Here,
\beq
	\rho(x)=\rho(x,x)=\int [dp]\tilde \rho(x,p).
\eeq
We must minimize this subject to the constraints
\beq
\tilde \rho\circ \tilde \rho(x,p)=\tilde \rho(x,p);\quad \tr\int
\tilde \rho(x,p)dx[dp]=m.
\eeq
We reiterate that although the problem has been formulated on the classical
phase space, no approximation has been made yet. All the complications are in
the multiplication law of the functions (hence in the quadratic constraint on
$\tilde \rho$).

Now we put in the separable ansatz (which satisfies the constraint automatically) and
expand in powers of $\hbar$.
We will have
\beq
	\tilde \rho(x,p)=\sum_{k=0}^{\infty}\tilde \rho_{(k)}(x,p){\hbar}^k
\eeq
Using the integral representation in terms of the resolvent symbol,
\beq
	\tilde \rho=\int_D{dE\over 2\pi i}\tilde r(E)
\eeq
we get
\beq
	\tilde \rho_{(k)}=\int_D {dE\over 2\pi i} \tilde r_{(k)}(E).
\eeq
These terms in the expansion of ${\tilde{\rho}}$ are distributions on the phase
space involving the delta function and its derivatives, although the $\tilde
r_{(k)}$  are ordinary functions.
In the same way, we have expansions for the number density,
\beq
	 \rho(x)=\sum_{k=0}^{\infty} \tilde \rho_{(k)}(x){\hbar}^k,
\eeq
\beq
\tilde \rho_{(k)}(x)=\int_D {dE\over 2\pi i} \int [dp]\tilde r_{(k)}(E,x,p)
\eeq
and the kinetic energy $ K=\tr  \int T(p)\tilde \rho(x,p) dx [dp]$,
\beq
	 K=\sum_{k=0}^{\infty} K_{(k)}{\hbar}^k.
\eeq
Also,
\beq
{\tilde q}_{(k)}=\int_D {dE\over 2\pi i} \tr\int dx [dp]\tilde
r_{(k)}(E,x,p)T(p).
\eeq
The direct energy can be written in terms of the density function $\rho(x)$.
The exchange integral is more complicated, being quadratic in $\tilde \rho$;
however, in most cases it is quite small and explicit calculation in higher
orders is not necessary.

It is now straightforward to calculate the Thomas--Fermi
energy to lowest order in the case of nonrelativistic quantum mechanics with a
potential $v$ that is diagonal in flavor space.  We get upon evaluating the integrals, 
(it is convenient to introduce  a new variable by $v(x)=-\phi^2(x)$),
\beqs
	\rho_{(0)}(x)&=&\int [dp]\Theta(-p^2+\phi(x))=\omega_n' {\phi^n(x)\over n},\cr
	K_{(0)}&=&\tr \int \omega_n'{\phi^{n+2}(x)\over n+2} dx.
\eeqs
Here
\beq
	\omega_n'={\omega_n\over (2\pi)^n}=2[{1\over 4\pi}]^{n\over 2}{1\over \Gamma({n\over 2})}
\eeq
is the area   of a sphere of unit radius in momentum space.
The exchange integral is,to lowest order,
\beq
	I_{(0)}=\half \int [dpdp']\Theta(\phi^2(x)-p^2)\Theta(\phi^2(x)-p'^2)\tilde
G(p-p').
\eeq
 With $\tilde G(p)={e^2\over p^2}$ as for the Coulomb interaction, we can evaluate this  more explicitly by introducing spherical polar coordinates in momentum space. We get 
\beq
	I_{(0)}=\half\alpha{\omega'_n\omega_{n-1}\over (2\pi)^{n}} C_n \int \phi^{2n-2}(x) dx 
\eeq
where
\beq
	C_n=\int_0^1dy\int_0^1 dy' (yy')^{(n-1)}\int_{0}^{\pi} d\theta
				{\sin^{n-2}\theta\over y^2+y'^2-2yy'\cos\theta}.
\eeq

This leads to
\beqs
E_{TF}(\phi)&=&\tr\int
[\omega_n'{\phi^{n+2}\over n+2}(x)-
{e^2\over 2}{\omega'_n\omega_{n-1}\over (2\pi)^{n}}C_n\phi^{2n-2}(x)
+ \omega_n'U(x){\phi^n(x)\over n}]dx+\cr
& & \half{\omega_n'^2\over n^2} \int G(x,y)\tr\phi(x)^n\tr\phi^n(y) dxdy.
\eeqs
In our expansion the exchange term appears in the lowest order. However, in
atomic physics it is as small as the terms involving derivatives of $\phi$, so
it is often ignored in the lowest order treatments.
Also, in many discussions, the energy is expressed as a function of the density
$\rho(x)$, but one can make the change of variable from the Fermi momentum
$\phi(x)$ to $\rho(x)$ easily.

Now we can vary this w.r.t. to \m{\phi} to get an integral equation that determines the ground state in this
approximation. Actually a more convenient variable to use is the mean field induced by this electron density:
in terms of it we get a differential equation instead. If the distribution is spherically symmetric, as for
an atom, this becomes a second order non-linear ordinary differential equation, the celebrated Thomas-Fermi
differential equation \cite{landaulifshitz}.
   
   Thus it is indeed possible to take the limit as \m{\hbar\to 0} on systems such as the atom and get a
   sensible approximation to the ground state. However, this leads  to  a 
   density function in  the
   classical phase space and not conventional classical mechanics. Moreover, it has to be derived through
   an intermediary that is a bona-fide classical mechanical system but of infinite dimensions.
   
   \section{ Atoms in the Limit of Large Dimension}

As another example of a neoclassical limit, again in atomic physics, we consider the limit of large spatial
dimension. This idea originates in an observation of Witten that in this limit the quantum fluctuations in
the rotation invariant quantities will become small. Let \m{r_{ai}} for \m{a=1,\cdots m} and \m{i=1,\cdots n}
be the positions of  \m{m}  electrons in an atom (or ion) of atomic number \m{Z}. Although the physically
interesting case is \m{n=3} we can, as a mathematical device, extend the system to \m{n} spatial dimensions.
The problem of determining the ground state becomes that of minimizing
\beq
\int \left[{\hbar^2\over 2\mu}{\pdr \psi^*\over \pdr r_{ai}}{\pdr \psi\over \pdr r_{ai}} 
 +\left(-\sum_{a}{Ze^2\over |r_a|}+ \sum_{1\leq a<b\leq m}{e^2\over |r_{a}-r_{b}|}\right)|\psi(r)|^2\right]\prod_{ai}dr_{ai}
\eeq 
subject to the condition that 
\beq
\int |\psi(r)|^2\prod_{ai}dr_{ai}=1.
\eeq
Here \m{\psi\in \Lambda^m\left(L^2(R^n,C^{N_f})\right)}.

Now, the hamiltonian is  invariant under the rotation group \m{O(n)\times U(N_f)}. We now take the 
limit as \m{n} and \m{N_f} tend to infinity, and recover a classical theory. When \m{N_f} is large, we can
assume that the wavefunction is completely anti-symmetric in the `spin' indices; the position dependent part
of the wavefunction is then symmetric. Indeed we can assume that this part  is rotation invariant,
\footnote{In the real world, the  ground state wavefunction is 
invariant under \m{O(n)} at least for the noble gases: all the shells are filled.} so that it 
depends only  the invariant quantities \m{q_{ab}={1\over n} r_{ai}r_{bi}}.

These inner products form a positive \m{m\times m} matrix.
 A complete set of \m{O(n)} invariants are given by
the remaining bilinears \m{\hat L_a^b={1\over 2n}[r_{aj},{\hbar\over i}{\pdr\over \pdr r_{aj}}]_+,
\hat P^{ab}=-{\hbar^2\over n}{\pdr^2\over \pdr r_{ai}\pdr r_{bi}}}. They form a representation 
 of the symplectic Lie algebra \m{Sp(2n)}:
\beqs
[q_{ab},q_{cd}]&=0=&[\hat P^{ab},\hat P^{cd}]\cr
[{\hat L}^a_b,{\hat L}^c_d]&=&{i\hbar \over n}\left(\delta_b^c{\hat L}_d^a-\delta_d^a{\hat L}_b^c\right)\cr
[{\hat L}^a_b,{\hat P}^{cd}]
&=&{i\hbar \over n}\left(\delta_b^c{\hat P}^{ad}+\delta_b^d{\hat P}^{ac}\right)\cr
[{\hat L}^a_b,q_{cd}]&=&-{i\hbar \over n}\left(\delta^a_c q_{bd}+\delta_d^a q_{bc}\right).
\eeqs

These commutators are proportional to \m{{\hbar \over n}}.
 Hence, there are two limits where the quantum fluctuations vanish: the 
 conventional classical limit where we let \m{\hbar\to 0}  keeping \m{n} fixed (at
 the value \m{3} for example), or the neo-classical limit where we let \m{n\to \infty} keeping \m{\hbar}
 fixed. In this neo-classical limit,
  the quantum observables tend to classical ones satisfying the  Poisson brackets of the 
  sympleclassical theory.ctic Lie algebra:
  \beqs
\{q_{ab},q_{cd}\}&=0=&\{ P^{ab}, P^{cd}\}\cr
\{{ L}^a_b,{ L}^c_d]&=&{\hbar}\left(\delta_b^c{ L}_d^a-\delta_d^a{ L}_b^c\right)\cr
\{{ L}^a_b,{ P}^{cd}\}
&=&{\hbar}\left(\delta_b^c{ P}^{ad}+\delta_b^d{ P}^{ac}\right)\cr
\{{ L}^a_b,q_{cd}\}&=&-{\hbar}\left(\delta^a_c q_{bd}+\delta_d^a q_{bc}\right).
\eeqs
These Poisson brackets will  determine the neo-classical equations of motion, once the hamiltonian is
determined.

 There are  some subtleties in determining the hamiltonian of this  neo-classical theory: there is a
 new term in the potential arising from the change of the measure of integration. (It is possible to
 interpret this as a kind of `Fischer information' while the measure determines a kind of `entropy'
 \cite{entropyinformation}. But we don't need this idea here.).
  Once the correct hamiltonian has been determined, this
   classical theory gives a relatively simple minimization problem for the ground state energy. Here we will consider only the static limit ( time
independent solution) that determines the ground state of theory. 

\subsection{The Change of Variables}

Let us return to  the variational problem of determining the ground state. This will reduce to the
minimization of an effective potential that depends only on \m{q_{ab}}. 
First of all, we need to determine the measure of integration \m{\mu(q)\prod_{c\leq d}dq_{cd}:=\mu(q)dq} 
determined by the change of variables \m{q_{ab}={1\over n} r_{ai}r_{bi}} on
the Lebesgue measure \m{\prod_{ai}dr_{ai}}.
This can be done by evaluating the following integral in two different ways:
\beq
Z(J)=\int_{q\geq 0} e^{-q_{ab}J^{ab}} \mu(q)\prod_{c\leq d}dq_{cd}=
\int e^{-{1\over n}r_{ai}r_{bi}J^{ab}}\prod_{cj}dr_{cj}.
\eeq
( Here, \m{J} is a positive matrix.)
On the r.h.s. we have a standard Gaussian integral yielding
\beq
\int_{q\geq 0} e^{-q_{ab}J^{ab}} \mu(q)\prod_{c\leq d}dq_{cd}=k(n,m)\left(\det J\right)^{-{n\over 2}}
\eeq
where \m{k(n,m)=\left(\pi n\right)^{-{nm\over 2}} } is independent of \m{J}. 

Thus \m{Z(J)} depends on \m{J} only through its determinant. It follows\footnote{ The space of positive
matrices is a homogenous space of the general linear group, since any such matrix can be mapped to the
identity by the transformation \m{q\mapsto SqS^T} with \m{J} transforming dually. 
The transformation law of \m{Z(J)} under this transformation completely determines that of \m{\mu(q)} as
well.} that \m{\mu(q)} can only depend on
\m{q} through \m{\det q}, prompting the ansatz \m{\mu(q)=\tilde k [\det q]^\nu}. To determine \m{\nu}  
we note that  under the transformation \m{q\to SqS^T}, the measure of integrtion transforms as 
\m{dq\mapsto [\det S]^{m+1} dq}. Thus
\beq
Z(J)=\int_{q\geq 0}e^{-\tr q(S^T J S)} [\det q]^\nu[\det S]^{2\nu+m+1}dq=
Z(S^TJS)[\det S]^{2\nu+m+1}.
\eeq
which determines \m{\nu={n-m-1\over 2}}.

Thus we have
\beq
||\psi||^2=\tilde k \int_{q\geq 0}|\psi(q)|^2 [\det q]^{n-m-1\over 2}dq.
\eeq
It is thus tempting to define a new wavefunction absorbing the determinant pf \m{q}:
\beq
\chi(q)=\surd\mu(q) \psi(q), \quad ||\psi||^2=
\int_{q\geq0} |\chi(q)|^2dq.
\eeq
This \m{\chi(q)} is a kind of `radial wavefunction'.

\subsection{The Effective Potential}

Now we must express  the hamiltonian in terms of this \m{\chi}. The only calculation we need  is for the
gradient of the wavefunction:
\beqs
\int {\hbar^2\over 2\mu n^2}{\pdr\psi^*\over \pdr r_{ai}}{\pdr \psi\over \pdr r_{ai}} \prod_{bj}dr_{bj}&=&
\int {\hbar^2\over 2\mu}g^{ab\ cd}
\mu^{-\half}{\pdr (\mu^{\half}\chi^*)\over \pdr q^{ab}}
\mu^{-\half}{\pdr (\mu^{\half}\chi)\over \pdr q^{cd}} dq\cr
&=&\int {\hbar^2\over 2\mu}g^{ab\ cd}\left[{\pdr \chi^*\over \pdr q^{ab}}+
{n-m-1\over 4}{\pdr \log\det q\over \pdr q^{ab}}\chi^*(q)\right]\cr
& & \left[{\pdr \chi\over \pdr q^{cd}}+
{n-m-1\over 4}{\pdr \log\det q\over \pdr q^{cd}}\chi(q)\right]dq
\eeqs
where 
\beq
g^{ab\ cd}=n^2{\pdr q^{ab}\over \pdr r_{ej}}{\pdr q^{cd}\over \pdr r_{ej}}=
\delta^{ac}q^{bd}+\delta^{ad}q^{bc}+\delta^{bc}q^{ad}+\delta^{bd}q^{ac}
\eeq
is an induced metric on the new configuration space. Moreover we know from elementary matrix theory that 
\beq
{\pdr \log\det q\over \pdr q^{ab}}=q^{-1}_{ab}.
\eeq
The terms prportional to  \m{\chi^*\chi } become a correction to the potential; the terms involving one 
derivative of the wavefunction combine to give  a total derivative that can be dropped. Those that involve
the square of the derivative of \m{\chi} become a new kinetic energy term. 

Thus the variational problem is now to minimize
\beq
\int_{q\geq 0}\left[{\hbar^2\over 2\mu n^2}g^{ab cd}{\pdr \chi^*\over \pdr q^{ab}}{\pdr \chi\over \pdr q^{cd}}
+\left(V_{eff}(q)+V(q)\right)|\chi(q)|^2\right]dq
\eeq
subject to the constraint
\beq
\int_{q\geq 0}|\chi(q)|^2 dq=1.
\eeq
Here,
\beq
V_{eff}={\hbar^2\over 2\mu}{(n-m-1)^2\over 4 n^2}\tr q^{-1}.
\eeq
Also, \m{U(q)} is the potential energy of the electron expressed in terms of the new variables:
\beq
U(q)=-\sum_{a=1}^m{Z\alpha\over \surd q^{aa}}+\sum_{1\leq a<b\leq n}{\alpha\over \surd[q^{aa}+q^{bb}-2q^{ab}]}
\eeq
where \m{\alpha={e^2\over \surd n}}.

We are now ready to take the limit as \m{n\to \infty}, holding \m{{e^2\over \surd n}=\alpha} ( not \m{e^2}
itself!) fixed. As expected, in the new variables,
the kinetic energy of the ground state wavefunction will be of order \m{{1\over
n^2}}. It is very important that there is now a  new term in the potential energy ( arising from the kinetic
energy of the old picture) which makes it bounded below:
\beq
V(q)={\hbar^2\over 8\mu}\tr q^{-1}+U(q).
\eeq
In the end the correction to the potential is quite simple!. 

The ground state energy in our neoclassical approximation is the minimum  this function over all positive
 \m{q}.
The condition for this is an algebraic equation for \m{q}.

 The case of a  hydrogenic ion is of course simplest: when \m{m=1}, \m{q} is just a positive number
 and there is no repulsive Coulomb interaction:
 \beq
 V(q)={\hbar^2\over 8\mu}q^{-1}-{Z\alpha\over \surd q}.
 \eeq
The minimum is  
\beq
-2\mu {Z^2\alpha^2\over \hbar^2}=-{2\over n}\mu{Z^2 e^4\over \hbar^2}.
\eeq
This is to be compared with the exact answer  ( for \m{n=3}) of \m{-\half \mu {Z^2 e^4\over \hbar^2}}. 
Thus we get roughly the correct answer: the relative error is about \m{1\over n}.  
It should be possible to improve on this by semi-classical methods.

More generally, it is reasonable to expect (but not guaranteed) that the minimum will respect the permutation
symmetry of the problem. Then we can put the ansatz that all the diagonal elements are equal 
(say, \m{q_{aa}=\rho^2,\forall a}) and that all the off-diagonal elements are also equal, 
(put \m{q_{ab}=\rho^2 u,\forall
a\neq b}). Then \m{|u|\leq 1} by Schwarz inequality. The potential  becomes in these new variables,
(it is convenient to choose a
kind of atomic units during such explicit calculations, \m{2\mu=\hbar=\alpha=1}):
\beq
V(\rho,u)={1\over 4\rho^2}f(u)-{1\over \rho}g(u).
\eeq 
Here, 
\beq
g(u)=mZ-{m(m-1)\over 2}{1\over \surd[2(1-u)]}.
\eeq

Moreover,
\beq
f(u)=\tr \tilde q^{-1},\quad  \tilde q=(1-u)+uC
\eeq
and \m{C} is the \m{m\times m} matrix all of whose matrix elements are equal to one. The spectrum of \m{C} is quite simple:
it has an eigenvalue equal to zero with degeneracy  \m{m-1} and the remaining eigenvalue is just \m{m}. Thus
we can determine the spectra of \m{\tilde q} and \m{\tilde q^{-1}} and hence its trace:
\beq
f(u)={m-1\over 1-u}+{1\over 1+(m-1)u}.
\eeq 

It is simple to minimize in \m{\rho} to reduce the problem to minimizing in \m{u} of 
\m{
-{g^2(u)\over f(u)}.
}
 If we change  variables yet again to 
\beq
v={1\over \surd[2(1-u)]}, \quad \half\leq v,
\eeq
our approximation of the ground state energy becomes the minimum of the rational function
\beq
\tilde V(v)=-\left[mZ-{m(m-1)\over 2}v\right]^2{2mv^2-(m-1)\over 2mv^2\left[2(m-1)v^2-(m-2)\right]}.
\eeq

This minimum can in fact be found in closed form as an algebraic 
 function of \m{m} and \m{Z}. But the formula (obtained by 
an algebraic computation program such as Mathematica) is quite complicated.
But this formula is fit very well\footnote{The fit is good to a relative error of  \m{0.01\%} over the
range \m{1\leq Z\leq 100}} in the case of a neutral atom (i.e., \m{m=Z}) by the
 polynomial
\beq
E(Z)=-\left[0.00152507+0.0871987 Z-0.920957 Z+1.83211 Z^2    \right]{2\mu e^4\over n\hbar^2}
\eeq
We have restored the original units to make comparison with other methods easier.

The point of this method is that it gives an exactly solvable
 and reasonably accurate picture for the ground state of the atom 
 without having
to deal with  complicated nonlinear differential equations.
The answers are reasonable considering the simplicity of the calculations.

   \section{ A Physicist's View of Modular Forms }
   
   Next we will consider an example from mathematics: the theory of modular forms. I don't claim to have
   solved any deep problem in this area ( of which there are many). But perhaps the point of view described
   will suggest new methods.
   
 \subsection{The Modular Group and its Subgroups}  
 We will give only a foretaste of the theory of modular forms. See
 Ref. \cite{apostol} for most of the proofs and precise statements of the
 results. Also see Ref. \cite{sarnak} for relations to other areas of mathematics and
 physics.

The group of two by two matrices with integer entries and determinant one is 
called \m{SL_2(Z)}; its  quotient  by the  center, 
\m{\Gamma(1)=SL_2(Z)/Z_2}, is  the {\Em modular group}. It is
conventional to denote elements of \m{\Gamma} as matrices
\m{\pmatrix{a&b\cr c&d}}, their pre-images in \m{SL_2(Z)}.

\m{\Gamma(1)} acts on the upper half of the complex plane \m{U} 
through the fractional linear transformations
\beq	z\mapsto {az+b\over cz+d}.
\eeq
A fundamental region for this action is,
\beq
	D=\{z||z|>1, |{\rm Re}\; z|<\half\}.
\eeq
This region is a spherical triangle with vertices at \m{i\infty,
\pm \half+{\surd 3\over 2}i}. The point is that translations can be
used to bring any point inside the strip \m{|{\rm Re}\; z|<\half};
and  under  inversion any point is equivalent to one  outside the unit
circle. 

The modular group is generated by 
\m{S:z\mapsto -{1\over z}} and \m{P=ST:z\mapsto -{1\over z+1}}. 
It is obvious that \m{S^2=1,P^3=1}. Indeed it can be shown that \m{\Gamma(1)=Z_2*Z_3}
 is the free product generated by these two elements-there are no other relations among these generators.

Many interesting groups appear as subgroups of the modular group. For example,
the commutator subgroup of \m{\Gamma(1)} is the free group on two
generators;
it is a normal subgroup of index\footnote{The {\Em index} \m{[G:H]} of a subgroup \m{H} of a group \m{G} is 
the number of elements in the coset \m{G/H}; alternatively, it is the number of copies of the fundamental 
region of \m{G} that is needed to form a fundamental region of \m{H}.} \m{6}. 
Thus the modular group is both non-abelian and infinite in an essential way: free groups are  the
ultimate examples of such groups.

 The
{\Em principal congruence subgroup \m{\Gamma(n)} of level \m{n}} 
consists of all
elements that are equal to the identity matrix modulo \m{n}.
Now we see why the modular group is called \m{\Gamma(1)}; its elements  are of the form
\m{\pmatrix{a&b\cr c&d}} with \m{a=d=1 \;{\rm mod}\;
n,c=b=0{\rm mod}\; n}. Any
subgroup \m{\Gamma} in between, \m{\Gamma(n)\subset \Gamma\subset
\Gamma(1)}, is called a {\Em congruence subgroup of level \m{n}}. 
The congruence subgroups
are all of finite index. It is possible to show by a counting argument \cite{apostol} that 
\m{[\Gamma(1):\Gamma(n)]=n^3\prod_{p|n}[1-p^{-3}].} 
Of particular importance \footnote{It is the modular forms of weight two with respect to this subgroup
that appear in the Shimura-Taniyama conjecture.} is the subgroup
\m{\Gamma_0(n)=\left\{\pmatrix{a&b\cr c&d}\in \Gamma(1),c=0\mod
n\right\}}.
 The index can be shown to be\cite{apostol} 
\m{[\Gamma(1):\Gamma_0(n)]=n\prod_{p|n}[1+p^{-1}]}.

\subsection{ Modular Forms}
A {\Em entire modular form} of integer {\Em weight} \m{k}  associated to a subgroup \m{\Gamma\subseteq
\Gamma(1)} is a
 holomorphic
 function
on the upper half plane (including the point at 
\footnote{ A function \m{f} is holomorphic
at \m{i\infty} if it has a convergent  Fourier expansion \m{f(z)=\sum_0^\infty
f_ne^{2\pi i n z}}. Moreover, \m{f(i\infty)=f_0}.}
\m{i\infty}) satisfying 
\beq
	(cz+d)^{-k}f\left({az+b\over cz+d}\right)=f(z), \;{\rm for}\; \pmatrix{a&b\cr c&d}\in \Gamma.
\eeq
It is called a {\Em cusp form} if \m{f(i\infty)=0}.

If the weight is even, we can think of a modular form as a
covariant tensor of order \m{k/2} (`form')  
 on \m{U/\Gamma}: the condition above is the
statement that \m{f(z)\left[dz\right]^{k\over 2}} is invariant
under \m{\Gamma}.

The homolomorphic sections of the 
canonical line bundle on \m{U} ( in this case the cotangent bundle) \m{\cal L}
 are entire functions \m{f(z)} on \m{U} such that \m{f(z)dz} is invariant under \m{\Gamma}. 
 Thus the modular forms of
weight \m{k}
are simply holomorphic sections of \m{{\cal L}^{k\over 2}}. If \m{k} is odd these correspond to some
`spinors' on \m{U/\Gamma}.

An example of a modular form\footnote{ If we don't specify  \m{\Gamma},
  we will be speaking of the modular group itself.}  of weight \m{2k} is the {\Em Eisenstein
series}
\beq
	G_{2k}(z)=\sum_{(m,n)\neq (0,0)}{1\over (m+nz)^{2k}}.
\eeq
It does not vanish at \m{i\infty}: \m{G_{2k}(i\infty)=2\zeta(2k)}.

The most famous cusp form  is 
\beq
	\Delta(z)=(2\pi)^{12}e^{i\pi z}\prod_{n=1}^\infty [1-e^{2\pi
inz}]^{12}.
\eeq
It is of weight \m{12}. It is nonzero everywhere except for a simple
zero at \m{i\infty}. It appears in Ramanujan's theory of partitions of numbers.
 If we expand the product of the twelveth root of  \m{\Delta} (which is called the Dedekind
 \m{\eta}-function) we can see that it is a generating function
for partitions. The partitions of large numbers is given by the asymptotic behavior as \m{{\rm Im}\; z \to 0}; this is an
essential singularity of the function so at first this looks hopeless. However, the modular invariance
relates the value of \m{\Delta} at \m{z} to its value at \m{-{1\over z}}; thus the behavior
at \m{i\infty} (which is trivial to determine) gives the behavior as \m{{\rm Im}\; z\to 0}. Hardy and Ramanujan turned
this rough stone of an idea  into an exquisite jewel (further polished by Rademacher), deriving an asymptotic formula
for partitions of large numbers.

 Any modular form is a periodic function hence can be expanded in a Fourier series. These Fourier
 coefficients are of great interest. An example
is the Ramanujan \m{\tau}-function, which are the Fourier coefficients of the modular form above,
\beq
	\Delta(z)=\sum_1^\infty \tau(n)e^{2\pi i nz}.
\eeq
A deep conjecture of Ramanujan  (proved eventually by Deligne following ideas of Grothendieck) was that 
\beq
	|\tau(p)|\leq 2p^{11\over 2}.
\eeq
In the theory of partitions, this inequality gives a  bound on 
the error term to the Hardy-Ramanujan asymptotic formula for partitions of large numbers. These error terms
seem to oscillate  erratically yet a  bound on their magnitude follows from the above inequality.

These erratic oscilations are related to yet  another interesting phenomenon:
if we define \m{2p^{11\over 2} \cos\theta(p)=\tau(p)}, the angles
\m{\theta(p)} seem to be distributed randomly according to the circular
 ensemble of random matrix theory. Indeed spectra of random matrices appear in many places in the theory of 
 modular forms and related Dirichlet series (see the recent books \cite{sarnak}). 
 We will seek   a clarification of this
 phenomenon using ideas from quantum mechanics in  the theory of Hecke operators.

Let \m{M_k} be the vector space of entire modular forms of weight \m{k} and
\m{{{\cal S}_k}} that of cusp forms. It is clear that \m{M_kM_l\subset
M_{k+l}}.
For \m{k\geq 12}, multiplication by
\m{\Delta} gives a linear map \m{M_{k-12}\to {{\cal S}_k}}. Moreover, \m{\dim {{\cal S}_k}=\dim M_{k}-1} since there is
just one condition on the Fourier coefficients of a cusp form: that
the  zeroth one vanishes. It is possible to  reduce the determination of the
dimension of \m{M_k} to small values of \m{k} using these facts; see \cite{apostol} for details. 

With our  definition of weight, there are no entire modular
forms of odd weight.
For \m{k=0} there is just one entire modular form, the constant. There
are none for \m{k=2}. For \m{k=4,6,8,10} the only entire modular forms
are multiples of the Eisenstein series.

For  \m{k\geq 12} and even, 
\beq
	\dim {{\cal S}_k}=\left\{\matrix{\left[k\over 12\right]\; {\rm if}\; k\neq 2
\mod 12\cr 
\left[k\over 12\right]-1\; {\rm if}\; k= 2
\mod 12}\right\}.
\eeq
Thus for large \m{k}, the dimension grows linearly with weight.A way of understanding this is that the
elements of \m{{{\cal S}_k}} are
holomorphic sections of the line bundle \m{{\cal L}^{k\over 2}}; as \m{k} grows this line bundle has greater 
Chern character allowing for more sections: it approaches a kind of classical limit.

 \subsection{ Modular Forms as Wavefunctions}

To a physicist,  the above theory of modular forms is very reminiscent of quantum mechanics.

We can regard the upper half plane as the phase space of some classical mechanical system. The 
symplectic form is the Poincar\`e form:
\beq
\omega={dx\wedge dy\over y^2}, \quad z=x+iy.
\eeq 
The modular group ( or one of its finite index subgroups) can be thought of a discrete gauge group, so that points related
by such a transformation represent the same classical state. A wavefunction would be a holomorphic function
on the upper half plane; more precisely it would be a holomorphic section of a line bundle \m{{\cal
L}^{k\over 2}} on \m{U/\Gamma}. Thus  \m{k\over 2} is analogous to the parameter \m{N} in our earlier
discussion of compact K\"ahler  manifolds. The base \m{U/\Gamma} is not usually a compact manifold because of
the  cusps ( points at infinity and points where the stability group is finite). Nevertheless \m{U/\Gamma} has
finite area hence most of the theory ought to generalize.

What is the dynamical system whose phase space is \m{U/\Gamma(1)}?. We can imagine it as a model of quantum
gravity in two dimensions. There are many such models that illustrate  various aspects of  gravity. Here,
we think of space-time as a torus. Our model of gravity is conformally invariant (not crazy since two is the
critical dimension for gravity). Thus the set of metrics modulo diffeomorphisms and conformal ( Weyl) 
transformations is the phase space of gravity. Using a diffeomorphism that is connected to the identity and a
Weyl transformation we can bring any metric to the form \m{ds^2=|d\theta_1+zd\theta_2|^2}, where
\m{0\leq \theta_1,\theta_2\leq 2\pi} are standard co-ordinates on the torus. Also we can choose 
 \m{{\rm Im}z>0}. Now if we also allow for diffeomorphisms that are not connected to the identity, which are
 \beq
 \pmatrix{\theta_1\cr \theta_2}\mapsto \pmatrix{a&b\cr c&d}\pmatrix{\theta_1\cr \theta_2}, \quad
 \pmatrix{a&b\cr c&d}\in SL_2(Z)
 \eeq
 then the true phase space would be \m{U/\Gamma(1)}: the action of \m{\Gamma(1)} on \m{z} is exactly the above
 fractional linear transformation.
 
 What would be the meaning of a gauge group that is only a subgroup of \m{\Gamma(1)}? We might have some
 additional geometric object on the torus that has to be invariant as well ( like a spin structure ) that would
 reduce the size of the gauge group.
 
 What would be the hamiltonian of our theory? A closed  cosmology like a  torus would at first not seem to
 have any meaningful time evolution. An asymptotically flat space-time would have a time at infinity with
 respect to which we can evolve its wavefunction. In closed universes in four dimensions, the Wheeler-DeWitt
 equation gives  a `time evolution' where the conformal factor of the metric itself is a kind of time
 variable. But we have given this up by postulating that continuous rescalings are part of the gauge group, so
 that the wavefunction is invariant under them.

  However we can still regard time evolution as a {\Em discrete} rescaling ( `expansion') of the
 universe. While  rescalings connected to the identity  are part of the conformal group, 
 there are certain discrete rescalings
 that for example double the size of the fundamental region. There are many different ways of rescaling such
 a region (e.g.,  double just one leg of the fundamental parallelogram) which individually violate modular
 invariance. Only by averaging over all of
 them would we recover modular invariance.
 
 What we describe above is an interpretation of the Hecke operators on modular forms:they are  rescalings averaged
 over the modular group. Because the  evolution is discrete we cannot find a
  generator for infinitesimal transformations. The closest we get to are the prime rescalings, which cannot
  be decomposed as compositions of others. They yield a
  family of commuting hermitean operators 
  which together play the role of the hamiltonian. In the next section we give a more detailed description of
  Hecke operators.
  
  The central problem of the Hecke theory of modular forms-determining the simultaneous eigenvectors of the Hecke
  operators- is just like the central problem in quantum mechanics- finding the eigenfunctions of the
  hamiltonian. Quantum mechanics suggests some strategies to attack this Hecke problem. The  limits of large
   weight or large level are like   neo-classical and classical limits. 
   For example, the number of linearly independent modular forms is \m{{\rm O}\left(\nu k\right)} in the limit
   of either large index \m{\nu} or large weight \m{k}. The  number of independent states of a quantum
   mechanical system with a two-dimensional phase space 
   is of order of the area of the phase space divided by \m{\hbar}. The fundamental region
   \m{D} of the modular group is not compact but still has finite area with respect to the Poincare metric:
   \beq
   A(D)=\int_{-\half}^\half dx\int_{\surd1-x^2}^\infty {dy\over y^2}={\pi\over 3}.
   \eeq
   The  fundamental region of a subgroup of index \m{\nu} is then just \m{\nu A(D)}.
   Since the number of linearly independent modular forms is \m{k\over 6} for large \m{k}, we see that the
   analogue of \m{\hbar} in our theory is essentially \m{6A(D)/k={2\pi\over k}}. For example for the subgroup 
    \m{\Gamma_0(p)} which has
   index \m{\sim p} (for prime \m{p}) there should be, according to this interpretation,
   \m{\sim pk} linearly independent modular forms. This is indeed known in the traditional theory of modular
   forms.

   Thus we get simpler classical analogues of
  the theory of modular forms in these limits; we can then hope to understand the general case by
  asymptotic  expansions in inverse powers of \m{k} or \m{n}. In the limit of large \m{n}, the subgroup
  \beq
  \Gamma_0(n)=\{\pmatrix{a&b\cr c&d}|c=0{\rm mod}\; n; ad-bc=1; a,b,c,d\in Z\}
  \eeq
  becomes essentially the group of translations
  \beq
  \pmatrix{1&c\cr 0&1}; c\in Z,\quad z\mapsto z+c.
  \eeq
   This is a huge simplification: the invariance group  becomes more `abelian'  as \m{n\to \infty}. In
   ordinary gauge theories ( such as Yang-Mills theories) the limit as the theory becomes abelian and the
   limit of small quantum corrections are intimately related. ( Perturbation theory is essentially the same
   as the loop expansion.) Thus one elementary strategy to understand modular forms is to study first this
   easy limiting case where modular forms reduce to periodic functions on the upper half plane. We will see
  then that this `perturbative' limit is also a `semi-classical limit'  of large \m{k}.
  
  The theory of modular forms should be viewed as a gauge theory with a non-abelian  and non-compact 
  gauge group- the modular group. It has all the essential features of the gauge theories of physics but in 
  a much simpler mathematical setting: the group is only countably infinite instead of being an  infinite
  dimensional Lie group. Thus there is no need for renormalization. By studying modular forms we are studying
  the  gauge principle  in its  purest form without contamination by the other complications of quantum field
  theories. The number theory of the last century bears witness to the claim that even this simplest of all
  non-abelian gauge theories is very deep: some of the deepest problems of number theory could be solved if
  we understood the spectrum of the Hecke operators.
  
  In this paper we develop only the analogue of lowest order pertrurbation theory ( `abelian approximation').
  Our other ideas on non-abelian gauge theories (`summing planar diagrams') also should have analogues here 
  and should lead
  to deep results in the future. I hope an  enterprising reader will take up this challenge.

\subsection{Hecke Operators}

We now return to the  exposition of the classic theory of modular forms due to Hecke. Hecke was motivated by
the work of Mordell who in turn was trying to understand the Ramanujan conjecture on the \m{\tau}-function.

A lattice on the complex plane is a set of points 
\beq
	w\sim w+r\omega_1+s\omega_2, \quad r,s\in Z;
\eeq
the fundamental region is a parallelogram with side \m{z}. A linear change of basis with integer
coefficients and determinant one\footnote{If the determinant is not one we change the area of the fundamental
domain.See below.} \m{ad-bc=1}, 
\beq
\pmatrix{\omega_1\cr\omega_2}\mapsto \pmatrix{a&b\cr c&d}\pmatrix{\omega_1\cr\omega_2}
\eeq
does not change the lattice.
By identifying the opposite sides of this
parallelogram we get a torus. By a rotating our co-ordinate system and choosing an appropriate unit of length
we can choose \m{\omega_2=1}. Also, we can reflect around this axis if needed to make \m{\omega_1} lie in the
upper half plane. Thus only the ratio \m{z={\omega_1\over \omega_2}} is needed to specify a lattice.
 A modular transformation 
\beq
	z\mapsto {az+b\over cz+d},\quad ad-bc=1
\eeq
is just the effect of a change of basis on this ratio. We call this lattice \m{L_z}.

A fractional linear transformation
\beq
	z\mapsto {az+b\over cz+d}
\eeq
with integer coefficients  will map \m{L_z} to a
sublattice of index\footnote{ This means that there are \m{n}
fundamental domains of \m{L_z} in one fundamental domain of the
sublattice.}  \m{n=ad-bc}. 
Each sublattice of index \m{n} corresponds to an orbit of \m{\Gamma(1)} on the 
set 
\beq
\Gamma_n=\left\{\pmatrix{a&b\cr c&d}|a,b,c,d\in Z, ad-bc=n\right\}
\eeq
since an action by \m{\Gamma(1)} would not have changed the original lattice \m{L_z}.
( For \m{n\neq 1}, \m{\Gamma_n} is not a group; 
\m{\Gamma_m\Gamma_n\subset \Gamma_{mn}}). 

We now define the action of the {\Em Hecke operator} \m{T(n)} on a modular 
form \m{f} as a sum  over all the sublattices of index\footnote{
We denote \m{h(z)={az+b\over cz+d}} for \m{h=\pmatrix{a&b\cr c&d} }}
 \m{n}:
\beq
	[T(n)f](z)=n^{{k\over 2}-1}\sum_{h\in \Gamma_n/\Gamma}f(h(z))
\left[dh(z)\over dz\right]^{ k/2}.
\eeq
Using the fact an action by \m{\Gamma(1)} merely permutes the terms of this sum ( the left action of
\m{\Gamma(1)} on the coset \m{\Gamma_n/\Gamma(1)}) we can show
 that  \m{T(n)f} is also a modular form of weight \m{k}. 

By a right action of \m{\Gamma(1)} we can bring any element of \m{\Gamma_n} to the upper triangular form
; actually  we can enumerate the elements of the coset \m{\Gamma_n/\Gamma(1)} by \m{\pmatrix{a&b\cr
0&d}} with \m{ad=n,b=0,1,\cdots d-1} . For proofs see \cite{apostol}.
A more explicit formula for the Hecke operator is thus,
\beq
	[T(n)f](z)={1\over n}\sum_{ad=n}a^k\sum_{b=0}^{d-1}
f\left({az+b\over d}\right).
\eeq
In particular, for prime \m{p},
\beq
[T(p)f](z)=p^{k-1}f(pz)+{1\over p}\sum_{b=0}^{p-1}f\left({z+b\over p}\right).
\eeq

In terms of Fourier coefficients:
\beq
f(z)=\sum_0^\infty f_me^{2\pi i mz},
[T(n)f](z)=\sum_0^\infty \gamma_n(m)e^{2\pi i mz},
\eeq
where,
\beq
\gamma_n(m)=\sum_{d|(n,m)}d^{k-1}f_{mn\over d^2}.
\eeq
In  particular\footnote{ \m{\delta(p|m)=1} if
 \m{ p}  divides \m{m} and zero otherwise. We use 
\m{\sum_{b=0}^{p-1}e^{2\pi im b\over p}=p\delta(p|m)} to derive this formula.},
\beq
	[T(p)f]_m=f_{pm}+\delta(p|m)p^{k-1}f_{m\over p}
\eeq

It follows then that \m{T(mn)=T(m)T(n)} if \m{m} and \m{n} are coprime. 
More generally we can show
\beq
	T(m)T(n)=\sum_{d|(m,n)}d^{k-1}T\left({mn\over d^2}\right).
\eeq
In particular, the Hecke operators commute with each other.
It is also useful that for prime powers we have a recursion relation,
\beq
	T(p^{r+1})=T(p)T(p^r)-p^{k-1}T(p^{r-1})
\eeq
which can be solved in terms of Tchebycheff polynomials\footnote{
The Tchebycheff polynomials are defined by 
\beq
	U_0(x)=1, U_1(x)=2x,U_{r+1}(x)=2xU_r(x)-U_{r-1}(x).
\eeq
}:
\beq
	T(p^r)=p^{r(k-1)\over 2}U_r\left(\half p^{k-1\over 2}T(p)\right).
\eeq
{\Em Thus the  \m{T(p)} for prime \m{p} determine all the \m{T(n)}.}

It is not difficult to establish an inner product on \m{{{\cal S}_k}}
with respect to which \m{T(n)} are hermitean:
\beq
	<f_1,f_2>=\int_{D} f_1^*(z)f_2(z) [{\rm Im }\;
z]^{k-2}d^2z,
\eeq
where,
\beq D=\{z=x+iy|-\half\leq x\leq \half,x^2+y^2\geq 1\}.
\eeq
The point here is that \m{y^{k}f_1^*(z)f_2(z)} is a modular invariant function, so that it can be integrated
after multiplying by the modular invariant volume form \m{dx\wedge dy\over y^2} to get a modular invariant 
quantity. Of course since each tile contributes the same amount we must restrict the integral to one
fundamental domain.

 Thus we have a set of commuting 
hermitean matrices, there is an orthogonal 
 basis of simultaneous eigenvectors, with real eigenvalues: the
 {\Em Hecke forms}. 

Suppose we have as simultaneous eigenvector a cuspform satisfying,
\beq
	T(n)f(z)=\lambda_nf(z),\quad \forall n.
\eeq

The convention is to normalize an eigenvector by setting the first
Fourier coefficient \m{f_1=1}.
Taking Fourier coefficients of both sides, we get \m{f_n=\lambda_n}:
{\Em the Fourier coefficients of a simultaneous eigenvector of the
Hecke operators are the eigenvalues}. It follows that these
coefficients satisfy a multiplicative identity
\beq
	f_mf_n=\sum_{d|(n,m)}d^{k-1}f_{mn\over d^2}.
\eeq

There is a  Dirichlet
series associated to any modular form
\beq
	f(z)=f_0+\sum_1^\infty f_n e^{2\pi i nz},\quad
\phi(s)=\sum_1^\infty {f_n\over n^s}
\eeq
or,
\beq
 \phi(s)={(2\pi)^s\over \Gamma(s)}\int_0^\infty
 y^{s-1}[f(iy)-f_0]dy.
\eeq

The modularity of \m{f} implies a functional equation\footnote{ To see this,
just take the Mellin transform of the condition for invariance under inversion:
\beq
f(-{1\over z})z^{k}=f(z).
\eeq
}
for \m{\phi}:
\beq
	(2\pi)^{-s}\Gamma(s)\phi(s)=(-1)^{k\over
2}(2\pi)^{s-k}\Gamma(k-s)\phi(k-s)
\eeq
and conversely. The multiplicative property of the coefficients of the
Hecke eigenforms yields a product formula:
\beq
	\phi(s)=\prod_p{1\over 1-f_pp^{-s}+p^{k-1}p^{-2s}}.
\eeq 
This is reminiscent of the Riemann zeta function \m{\zeta(s)}. Indeed the theta
function, of which \m{\zeta(s)} is the Mellin transform, is a modular
form of a congruence subgroup of level 2.

It is of much interest to understand the behavior of the eigenvalues
of the Hecke operators. They have been related to the zeros of the zeta function of algebraic varieties
 over finite fields (Eichler, Sato, Deligne).
 There are 
 many deep conjectures about the
behavior of the eigenvalues \m{\lambda(n)} for large \m{n}. The simplest case is when the dimension of the
space of cusp
forms is one: when \m{k=12} the only cusp form is the function \m{\Delta(z)} we introduced earlier. In this
case the Hecke operators are  \m{1\times 1} matrices: just numbers. From the above it is clear that these 
numbers are just the Fourier coefficients of the function \m{\Delta(z)}. In other words, for the case
\m{k=12}, the Hecke operators reduce to the Ramanujan \m{\tau}-function: \m{T(n)=\tau(n)}. In fact Hecke
discovered these operators  by generalizing some ideas of Mordell on the modular form \m{\Delta(z)} to the
case of higher weight.

From our earlier discussion, we are led to consider the limit of large level where the invariance group
becomes abelian. We now present a simple analogue of the Hecke problem for this case of periodic functions.

\subsection{Hecke Operators on Periodic Functions}

Any modular form is periodic so we can expand it in a Fourier series \m{f(z)=\sum_{n=1}f_ne_n(z)}. Thus it is
tempting to think of the space of modular forms as a subspace of the space of periodic functions \m{V}.
To make this idea precise, we would like to have a norm on \m{V}.

The inner product on modular forms given above can be written as 
\beq
<f,\tilde f>=\int_{-\half}^\half dx\int_{x^2+y^2\geq 1}dy y^{k-2}f^*(z)\tilde f(z)=\sum_{m,n=1}^\infty f_n^*\tilde f_mg_{nm}
\eeq
where
\beqs
g_{nm}=<e_n,e_m>&=&\int_{-\half}^\half dx e^{2\pi i(m-n)x} \int_0^\infty  
y^{k-2}e^{-2\pi(m+n)y}\cr
& & \theta\left(y\leq \surd(1-x^2)\right) dy
\eeqs
is a positive sesquilinear  form. 
We can extend the integral to the fundamental region of the translation group to get an inner product on
\m{V}:
\beq
(\psi,\tilde \psi)=\int_{-\half}^\half dx\int_0^\infty y^{k-2}\psi^*\tilde\psi dy=\sum_{n,m=1}^\infty
\psi^*_n\tilde\psi_m h_{nm}.
\eeq 
In terms of the basis \m{e_m}, we have   \m{h_{nm}}  as a  diagonal sequilinear form
\beq
h_{nm}=(e_n,e_m)=(k-2)![4\pi n]^{1-k}\delta_{n,m}.
\eeq

The Hecke operators are given by quite simple formulae in terms of the Fourier components. We can work out
easily their dual action on the basis \m{e_m(z)}:
\beq
	T(p)e_m(z)=p^{k-1}e_{mp}(z)+\delta(p|m)e_{m\over p}(z)
\eeq
for prime \m{p}. This can then be used extend to them as  operators on \m{V}.
 The spectral
problem for \m{T(p)} in the infinite dimensional space \m{V} is much simpler than its counterpart in \m{{{\cal S}_k}}.
We will solve this simpler problem and see that it has close connections to the theory of random matrices.

How will we recover the Hecke operators on modular forms? We could try thinking  of modular forms as a subspace of
the space of periodic functions. But under the above inner product on \m{V} they will not be square
integrable. The reason is precisely modular invariance: each fundamental region contributes an equal amount
to the integral, and the region \m{-\half\leq x\leq \half} contains an infinite number of such regions.
Thus, the expansion of a modular form in the basis \m{e_m(z)} is not convergent in the above norm \m{(.,)}. On the
other hand, the integral for  \m{<.,.>} corresponding to the sesqilinear form \m{g_{nm}} also can be extended
to an inner product on \m{V}. It is convergent on modular forms but is a  degenerate  sesquilinear form in
\m{V}: the integral is restricted to the region \m{x^2+y^2\geq 1}. We can quotient \m{V} by the null space of
\m{g_{nm}} to get  a finite dimensional space \m{\tilde{\cal S}_k}
that is a `gauge fixed' version of \m{{\cal S}_k}. 
That is, instead of thinking of modular forms as a subspace of \m{V},
we think of them as a quotient of \m{V}
by the null space of \m{g_{nm}}.
The `gauge fixing'   amounts to the choice of one
particular fundamental region among the infinite number as the domain of the integral. Thus the 
 spectral problem of the Hecke operators on modular forms can be replaced by that on the space \m{\tilde{\cal
 S}_k}.

As \m{k} grows the dimension of \m{{\cal S}_k} grows linearly with \m{k}. We will see that in a certain sense
the two inner products \m{g_{nm}} and \m{h_{nm}} approach each other. Thus the discrete eigenvalues 
 of \m{T(p)}  are so close together as \m{k\to \infty} that they merge to the form the continuous spectrum of
 \m{T(p)} on \m{V}.

\subsection{Toeplitz Operators}

We now solve the spectral problem for the Hecke operators on periodic functions.
It will be more transparent to transform to  the orthonormal basis 
\m{|m>=\left[(4\pi m)^{1-k}(k-2)!\right]^{-\half}e_m} we have
\beq
	T(p)|m>=p^{k-1\over 2}\left[|mp>+\delta(p|m)|{m\over p}>\right].
\eeq

There is then a simple description in terms of Toeplitz operators. 
The operators 
\beq
	A^{\dag}(p)|m>=|pm>, A(p)|m>=\delta(p|m)|{m\over p}>
\eeq
are adjoints of each other and satisfy
\beq
	A(p)A^{\dag}(p)=1,A(p)A(p')=A(p)A(p'),A(p)A^\dag(p')=A^\dag(p')A(p)
\eeq
for \m{p\neq p'}.
That is, they form a commuting set of Toeplitz operators labelled by
the prime  numbers. Being isometries \m{|A|=|A^\dag|=1}. It follows
easily that \m{T(p)} are hermitean and that 
\beq
	|T(p)|\leq 2p^{k-1\over 2}.
\eeq
The analogue of this inequality on the space of modular invariant
functions 
(rather than periodic functions) is a much deeper statement.

\subsection{ Connection to  Random Matrices}

The simultaneous eigenfunctions of the \m{T(p)} can now be obtained in
terms of Tchebychev polynomials. The spectrum is connected with the
Wigner distribution for random matrices. 

It is enough to study each \m{T(p)} separately: the theory `localizes'
completely. To see this, represent each number \m{m} in terms of its
prime decomposition:
\beq
	m=\prod_p p^{\nu_p}.
\eeq
The product is over the set of all primes; \m{\nu_p=0,1,\cdots} with
only a finite number of them being non-zero. Then 
\beq
	A^\dag(p)|\nu_2,\nu_3,\cdots>=|\nu_2,\cdots \nu_p+1,\cdots>,
\eeq
\beq
A(p)|\nu_2,\nu_3,\cdots>=\delta(\nu_p\neq 0)|\nu_2,\cdots\nu_p-1,\cdots>.
\eeq
Thus \m{A(p), A^\dag(p)} act only on the \m{p}-th entry.

The Toeplitz algebra is the associative algebra generated by a pair of
elements satisfying the relation
\beq
	AA^\dag=1.
\eeq
The standard representation in terms of an orthonormal basis
\m{|\nu>,\nu=0,1\cdots} is
\beq
	A^\dag|\nu>=|\nu+1>,\quad A|\nu>=\delta(\nu\neq 0)|\nu-1>.
\eeq
This is precisely the representation that we have. 

Voiculescu \cite{randommatrix} has found a remarkable connection between the theory of
random matrices and the Toeplitz algebra. 
 Given any polynomial \m{f:R\to R},
define
\beq
	<f>_N={\int \tr f(X)e^{-\half \tr X^2} dX\over \int e^{-\tr X^2} dX}
\eeq
the integral being over over all hermitean \m{N\times N} matrices.
 Thus \m{X} is a
hermitean matrix whose matrix elements are independent random
variables. Then, Voiculescu shows that 
\beq
	\lim_{N\to \infty} <f>_N=<0|f\left(A+A^\dag\right)|0>.
\eeq
 There is a probability distribution on \m{R},
the Wigner semi-circle distribution, such that 
\beq
	\lim_{N\to \infty}<f>_N=\int_R f(x) \rho(x)dx.
\eeq	
Explicitly,
\beq
	\rho(x)=\theta(|x|<2){1\over 2\pi}\surd[4-x^2].
\eeq

Thus we see that the Hecke  operators \m{T(p)} on periodic functions are
hermitean operators whose spectrum is  the interval
\m{[-2p^{k-1\over 2}, 2p^{k-2\over 2}]}. The (generalized)
eigenfunctions are given by Tchebycheff polynomials.  The Wigner distribution gives the spectral density.
Thus each \m{T(p)} behaves like a hermitean random matrix; the different Hecke operators for different 
prime \m{p}  commute with each other, so they are not `free' in the sense of Voiculescu; instead they are
statistically independent in the more conventional sense.

\subsection{ The limit of Large Weight}

Recall  that the main difference between the exactly solvable model above and the
the theory of modular forms is that we replaced the sesquilinear form \m{g_{nm}} by the simpler one
\m{h_{nm}}. We now show that in the limit of large weight \m{k} this is a small correction so that what we
obtained above is the asymptotic behavior as \m{k\to \infty}.

Note that we can split 
\beq
h_{nm}=g_{nm}+q_{nm}
\eeq
where \m{k} is the integral over the complimentary region
\beq
q_{nm}=\int_{-\half}^\half dx e^{2\pi i (m-n)x}\int_0^{\surd(1-x^2)} y^{k-2}e^{-2\pi(m+n)y}dy.
\eeq
If we rewrite this in the orthonormal basis of \m{h_{nm}}, we will get
\beq
\delta_{nm}=\tilde g_{nm}+\tilde q_{nm}
\eeq
where \m{\tilde q_{nm}=q_{nm}/{\surd(h_{nn}h_{mm})}} etc . We will show that \m{|\tilde q_{nm}|} tends to zero
as \m{k\to \infty}.

Now,evaluating the \m{y}-integral,
\beqs
\tilde q_{nm}&=&\left[{{m+n\over 2}\over \surd(mn)}\right]^{1-k}\cr
& & \int_{-\half}^\half dx e^{2\pi i (m-n)x}{1\over (k-2)!}\Gamma\bigg(k-1,2\pi(m+n)\surd(1-x^2)\bigg)
\eeqs
where the incomplete Gamma function is defined by 
\beq
\Gamma(s,u)=\int_0^u t^{s-1}e^{-t}dt.
\eeq
 
 We will study this limit as \m{k\to \infty} keeping \m{m,n} fixed\footnote{ We should really be estimating
 the operator norm  of \m{{\tilde q}}. I hope that the arguments here will motivate  a more rigorous
 analysis. 
 }. Now recall that as \m{s-1>u}, the maximum value of the integrand is attained
 at its upper limit in this case, so that \m{\Gamma(s,u)< u^se^{-u}}. Then
 \beqs
 |\tilde q_{nm}|&\leq& \left[{{m+n\over 2}\over \surd(mn)}\right]^{1-k} \int_{-\half}^\half 
 {1\over (k-2)!}\cr
 & & \left[2\pi(m+n)\surd(1-x^2)\right]^{k-1}e^{-2\pi(m+n)\surd(1-x^2)}dx.
 \eeqs  
Again, replacing the integrand by its largest value ( which is attained at \m{x=0}), we get
\beq
|\tilde q_{nm}|\leq {\left[4\pi\surd(mn)\right]^{k-1}\over (k-2)!}.
\eeq
The growth of the factorial beats the exponential growth for fixed \m{m} and \m{n}.

Thus we can see why the distribution of eigenvalues of Hecke operators resemble those of random matrices by
our semi-classical approximation method. Moreover we see why the spectrum of \m{T(p)} is in the interval 
\m{[-2p^{k-1\over 2},2p^{k-1\over 2}]}.
\subsection{The Poisson Algebra of Modular invariant functions}  

We should expect that in the limit of large weight, the theory of modular forms is well-approximated by a
classical theory. More precisely the algebra of matrices on \m{{{\cal S}_k}} should tend to the Poisson algebra of
functions on \m{U/\Gamma(1)}. In particular there will be functions on the upper half plane which are
classical  approximations to the Hecke operators. These Hecke functions will have vanishing Poisson brackets
relative to each other. Their range will give a classical approximation to the Hecke eigenvalue problem.
This is another, (manifestly gauge invariant)  of studying the limit of large weight.

The upper half plane is  a symplectic manifold with symplectic form
\beq
	\omega={dx\wedge dy\over y^2}.
\eeq
This means that \m{x} and \m{{1\over y}} are canonical conjugates:
\beq
	\{y^{-1},x\}=1.
\eeq
The set of functions on the upper halfplane form a Poisson algebra
with the bracket
\beq	
	\{u,v\}=y^2\left( {\pdr u\over \pdr x}{\pdr v\over \pdr
y}-{\pdr u\over \pdr y}{\pdr v\over \pdr x}\right).
\eeq

The modular transformations
\beq
	x\mapsto x+1, y\mapsto y
\eeq
and
\beq
	x\to -{x\over x^2+y^2}, y\mapsto {y\over x^2+y^2}
\eeq
are canonical transformations. Thus the space of modular invariant
functions is a sub-Poisson algebra. The quotient of this by its center is the 
algebra of `gauge invariant observables' if  we 
regard the modular group as a `gauge group'. 
We can construct such observables from smooth functions of the upper
half plane (vanishing sufficiently fast at infinity) by averaging over orbits. The Maas forms provide
 nice examples of such `observables'.

In the limit of large \m{k}, the Hecke operators should tend to certain modular invariant functions
( which are not holomorphic) that have zero Poisson brackets relative to each other.
The range of these functions is the large \m{k}-limit of the Hecke spectrum.
 We should also be able to derive a systematic semi-classical expansion  in powers of \m{1\over k}.
  But this paper is getting long already; I hope to return to
these questions in a later publication.

\centerline{\bf Acknowledgement}

I thank Teoman Turgut and Edwin Langmann warmly for   reading through
 various versions of this paper. Also, Ersan Demilrap
brought references \cite{hershbach,hartreefock, densityfunctional} to my attention.
Thanks are due also to  A. Agarwal,  L. Akant  and  G. Krishnaswami for many  discussions.
The gracious hospitality of the Erwin Schr\"odinger Institute (Vienna) and the Feza G\"ursey 
Institute (Istanbul), where this paper was written,  is also acknowledged.

\end{document}